\documentclass[aps,pra,twocolumn,showpacs,preprintnumbers,amsmath,amssymb,footinbib]{revtex4}
\usepackage{graphicx,epsfig}
\usepackage{bm}
\usepackage{dcolumn}
\usepackage{xcolor}
\usepackage[breaklinks=true,colorlinks,citecolor=blue,linkcolor=blue,urlcolor=blue]{hyperref}

\def\noi{\noindent}
\def\bc{\begin{center}}
\def\ec{\end{center}}
\newcommand{\bea}{\begin{equation}}
\newcommand{\eea}{\end{equation}\noi}
\newcommand{\ber}{\begin{eqnarray}}
\newcommand{\eer}{\end{eqnarray}\noi}
\usepackage{calrsfs}
\DeclareMathAlphabet{\pazocal}{OMS}{zplm}{m}{n}
\newcommand{\Anna}{\mathcal{A}}

\usepackage{amsfonts}
\usepackage{calligra}
\usepackage{amssymb,amsmath}
\usepackage{caption}

\usepackage{graphicx}
\usepackage{caption}
\usepackage{comment}

\begin{document}

\title{A study on Stokes-Brinkman dimensionless model for flow in porous media}

\author{Anna Caroline Felix Santos de Jesus $^{1}$}\email{annacfelixs@gmail.com, carolinefelix@usp.br}
\affiliation{$^{1}$ Laboratory of Applied Mathematics and Scientific Computing (LMACC),
University of Sao Paulo,
Sao Paulo, Brazil}

\date{\today}

\begin{abstract}

In this work we propose a non-dimensional formulation for the Stokes-Brinkman model for the flow in porous media. We studied the effect of the dimensionless number found, which will be denoted by $ \Anna $
and named as Anna's number. This parameter is important to observe the transition from the Darcy regime to the pure Stokes regime.
\end{abstract}


\maketitle
 

\section{Introduction}

The Brinkman model is applied to media having a transition structure between a porous medium and another medium in which the inertial forces can be ignored. H. C. Brinkman \cite{Brinkman1949}, made an important observation in 1949 that he added the viscous term $\mu '\nabla^2$ to the Darcy equation resulting in the following model for incompressible flow
\begin{align}
\mu' \nabla^2 \textbf{u} + \mu \textbf{K}^{-1} \textbf{u} + \nabla p &= \textbf{f}, \quad \quad  \mathrm{in} \quad \Omega, \label{eq:brinkman} \\
\nabla \cdot \textbf{u} &= 0, \quad \quad  \mathrm{in} \quad \Omega, \label{eq:conservacao-massa} \\
\textbf{u} &= \textbf{g} , \quad \quad  \mathrm{on} \quad \partial \Omega \label{eq:condicao-contorno}.
\end{align}
where $\Omega \subset \mathbb{R}^d$, (d = 2 or 3) is open and limited. In this model, $p$ is the fluid pressure, $\textbf{u}$ is the velocity, $ \textbf{K}$ is the tensor permeability of the porous medium, and can be highly heterogeneous, $ \mu $ is the viscosity of fluid, $ \mu'$ is the so-called effective viscosity of the fluid, and f denotes the term force.
Choosing the parameters appropriately, one can study the behavior of the model for boundary cases, where: for $ K_{ij} \to \infty $, we obtain the Stokes flow and for $ K_{ij} \to 0 $, we have the Darcy flow. This characteristic is ideal for problems in carbonate rocks in the presence of fractures and/or cavities, and also for flow in highly porous filters and applications in biomedicine. Further, in the literature, it is often assumed that $ \mu = \mu'$.

In order to analyze the effects of the parameters in the solution, equations \eqref{eq:brinkman} - \eqref{eq:condicao-contorno} will be studied in their dimensionless form. Depending on the adopted parameters, the dimensionlessness constant obtained, which will be called the Anna number, characterizes the flow regime, indicating if the flow is dominated by the model of Darcy, Brinkman or pure Stokes.

\section{\label{sec:one} Dimensional Stokes-Brinkman Model}

Disregarding gravitational effects, to dimensionalize the Stokes-Brinkman equation, in the stationary case, we define the following dimensionless variables

\begin{align} \label{grandezas-adimensionais-brinkman}
\textbf{x}^* = \frac{\textbf{x}}{\Tilde{L}}, \quad
\textbf{u}^* = \frac{\textbf{u}}{\Tilde{U}}, \quad
p^* = \frac{p}{\Tilde{p}}, \quad \textbf{K}^* = \frac{1}{K_{max}} \textbf{K} 
\end{align}
and the dimensionless operator
\begin{align}\label{operador-adimensional}
\quad  \nabla^* = \Tilde{L} \nabla,
\end{align}
with
\begin{align}  \label{grandezas-adimensionais-brinkman-2}
\Tilde{p} = \frac{\Tilde{L} \Tilde{U} \mu}{K_{max}}.
\end{align}
Here $\Tilde{L}$ and $\Tilde{U} $ are respectively reference values of length and velocity and $K_{max}$ is the largest input value of the permeability tensor.

Substituting the dimensionless values indicated by $^*$ into \eqref{grandezas-adimensionais-brinkman} and \eqref {operador-adimensional} in the equation \eqref{eq:brinkman}, follows

\begin{align*}
-\mu'   \frac{\Tilde{U}^2}{\Tilde{L}^2}  \nabla^{*^2}  \textbf{u}^*
+ \mu  \frac{\Tilde{U}}{K_{max}}   \textbf{K}^{*^{-1}}    \textbf{u}^* +    \frac{\mu \Tilde{L}\Tilde{U}}{K_{max}} + \nabla^* p^*   = 0 
\end{align*}
Multiplying both members of the last equation by $ \frac{K_{max}} {\mu \Tilde{U}}$ we get
\begin{align}  
- \bigg(\frac{\mu'}{\mu} \frac{K_{max}}{\Tilde{L}^2} \bigg) \nabla^{*^2}  \textbf{u}^* +  \textbf{K}^{*^{-1}}  
\textbf{u}^* + \nabla^{*}  p^* = 0, \nonumber \\
- \Anna \nabla^{*^2} \textbf{u}^* +  \textbf{K}^{*^{-1}}   \textbf{u}^* + \nabla^* p^* = 0, \label{eq:adimensional-brinkman}
\end{align} 
where we denote by $\Anna = \frac{\mu'}{\mu} \frac{K_{max}}{\Tilde{L}^2} = \frac{\mu'}{\mu} Da $. This dimensionless number generalizes the Darcy Number (Da) for cases where the effective viscosity and viscosity of the fluid are not of the same order.
On the other hand, for the mass conservation equation \eqref{eq:conservacao-massa} it is very easy to verify that

\begin{align*}
\nabla ^* \cdot \textbf{u}^* = 0. 
\end{align*}


Finally, eliminating the asterisks from the equations and using the definition of the number Anna, we have the Stokes-Brinkman equation in the dimensionless form,

\begin{align}
- \Anna \nabla^2 \textbf{u} +  \textbf{K}^{-1} \textbf{u} + \nabla p & = \textbf{0} \quad \text{in} \quad \Omega,  \label{eq:bb-adimensional} \\
\nabla \cdot \textbf{u} & = 0 \quad \text{in} \quad \Omega, \\
\textbf{u} & = \Tilde{\textbf{g}} \quad \text{on} \quad \partial \Omega.  \label{eq:cc-adimensional}
\end{align}

It is evident that the Stokes term becomes significant only when one works with non-negligible Darcy numbers.
In fact, we observe that when $\Anna $ is small the equation \eqref{eq:adimensional-brinkman} tends to the Darcy equation, and when it is significant, the Brinkman equation distances itself from Darcy and therefore the number dimensionless $\Anna$ is fundamental to delimit numerically when the use of the Brinkman equation for flow is significant.
However, even if values not negligibe for the Darcy number ($Da$)
are considered, some authors still argue from which porosity there would be a physical coherence in modeling the flow with the Brinkman equation. According to \cite{artigo:nield2013} Brinkman's formulation for porous media ceases to be consistent for porosities below 0.6.

\section{\label{sec:two} The finite volume method for the model in dimensionless form} 
The finite volume method is the most widespread in reservoir simulation because it also meets the principle of mass conservation at the local level, in the present work we adopt this method in the process of discretization of the differential equations.

The application of the finite volume method basically consists of subdividing the problem domain into contiguous control volumes and integrating the differential equation into each control volume.

The discretization of the Brinkman model described by the equations \eqref{eq:brinkman} - \eqref{eq:condicao-contorno} is performed in a staggered grid, where the pressure unknowns are computed in the centers of the control volumes, while the velocities are approximated in the faces of the volumes.

The equations \eqref{eq:bb-adimensional} - \eqref {eq:cc-adimensional}, once discretized through the formulation of finite volumes, lead to an algebraic system as follows

\begin{align}\label{sistema-monolitico}
\begin{bmatrix}
\textbf{A} & \textbf{G} \\ 
\textbf{D} & 0
\end{bmatrix}  \begin{bmatrix}
\textbf{U}  \\ 
\textbf{P}
\end{bmatrix} =\begin{bmatrix}
\textbf{0}  \\ 
0
\end{bmatrix}. 
\end{align}
onde 
\begin{align*}
\textbf{A} = -\Anna \textbf{L} + \textbf{K}^{-1} \mathbb{I},
\end{align*}
$ \mathbb{I} $ is the identity matrix; $ \textbf{L} $ is the Laplacian operator; $ \textbf{G} $ is the gradient operator and $ \textbf{D} $ is the divergence operator. In the vector $ \textbf{U} $ was added the known values for the velocity.

We call this a coupled, or monolithic, method to solve the algebraic system \eqref{sistema-monolitico} by direct or iterative methods, without decomposing the vector of unknowns into pressure and velocity unknowns.

The implementation was performed with the MATLAB software and the linear system solved with the backslash command. In MATLAB, this operator encompasses several algorithms to manipulate different types of input arrays. Thus, the input matrix is diagnosed and an execution path is selected according to its characteristics.

\section{\label{sec:tres} Numerical Experiments}

There are still on going discussions about the validity and applications of Brinkman equations as a flow model in porous media with higher porosity values. However, the Stokes-Brinkman equation system is a convenient model of flows in highly heterogeneous porous media with random distribution of obstacles, channels, cavities, and other geophysical characteristics \cite{KANSCHAT2017174}.

According to \cite{artigo:christie} geological models of reservoirs may present problems of the order of $ 10^8 $ incognitos, which gives rise to large linear system. In addition, it is known in the literature that the fact that they represent complex physical phenomena, generally results in poorly conditioned systems from the point of view of linear algebra. This is especially true when the permeability contrast $K_{max}/ K_{min}$ is very high, that is, with great variability along the space that entails in the deterioration of the convergence of iterative numerical methods that propose to solve it.
Given the above and knowing that the matrix of the coefficients is sparse and not symmetrical, the iterative method used was the Generalized Minimal Residual Method - GMRES.

\subsection{Heterogeneous Anisotropic Case}

Initially we consider the anisotropic heterogeneous case, that is, $K_{xx}(x, y) \neq K_{yy}(x, y) $ whose heterogeneous high contrast permeability field, contrast $\sim 10^5$ in the x-direction and contrast $\sim 10^5$ in the y-direction
for a two-dimensional domain $\Omega = (0,1) \times (0,1) $. The boundary condition used is $ \textbf{g} = (1.0) $.

The coefficient matrix is not symmetric and has dimension $1240^2$; the tolerance adopted is the standard, 
namely $tol = 10^{-6}$ and maximum permissible iterations is $maxit = 1240 $;

We consider the trivial case in which $\mu' = \mu$ and that therefore number of Anna coincides with the number of Darcy. Note that if we fix $\Tilde{L} = 1$ the Darcy number increases if $K_{max}$ increases, which is equivalent to increasing the permeability contrast in both the x-direction and the y-direction.

Note that as the Darcy number increases the condition number of the matrix it also increases, that is, the system becomes increasingly poorly conditioned, as shown in the \autoref{tabela:comportamento_gmres}. However, we see an improvement in convergence when the system enters the Stokes regime, that is, the permeability field is ignored and this leads to an improvement in convergence.

\begin{table}[h!]
	\centering
	\begin{tabular}{cccc}
		\hline
		Da & $\kappa(A)$ & Iterações      \\
		\hline
		$10^{-5}$    & $5.76110\cdot 10^8$        & 1154    \\
		$10^{-4}$    & $5.76191\cdot 10^8$        & 1154               \\
		$10^{-3}$    & $5.76986 \cdot 10^8$        & 1154                 \\
		$10^{-2}$    & $5.84419\cdot 10^8$        & 1149                 \\
		$10^{-1}$    & $6.43460\cdot 10^8$        & 997                  \\
		$10^0$            & $1.03837\cdot 10^9$        & 821                  \\
		$10^1$           & $4.38194 \cdot 10^9$        & 607                  \\
		$10^2$       & $6.35553 \cdot 10^{10}$        & 341                 \\
		$10^3$       & $3.50124\cdot 10^{12}$        & 123                  \\
		$10^4$       & $3.22691 \cdot 10^{14}$        & 69                   \\
		$10^5$       & $3.19956 \cdot 10^{16}$        & 68                   \\
		\hline
	\end{tabular}
	\caption{Behavior of the GMRES as a function of the condition number of the system matrix
	} \label{tabela:comportamento_gmres}
\end{table}

An explanation for this phenomenon is that in the specific case of the GMRES, the condition number of the matrix nothing interferes to draw conclusions about the convergence of the system. The distribution of eigenvalues is what dictates the conclusions. The more distant from the origin the eigenvalues are distributed, the better the convergence of the method. And this was observed experimentally.
The regime in which the transition from the Darcy model to the pure Stokes model occurs, occurs when Darcy's number approaches unity.


\end{document}